\documentclass[12pt,preprint]{elsarticle}

\usepackage{epsfig}
\usepackage{amsmath,amssymb,amsthm}
\usepackage{graphicx}
\usepackage{psfrag}
\usepackage{bm}

\journal{Phys. Lett. A}

\begin{document}
\begin{frontmatter}

\title{Fisher information, nonclassicality and quantum revivals}

\author{Elvira Romera}
\address{
Instituto Carlos I de F{\'\i}sica Te\'orica y
Computacional, Universidad de Granada, Fuentenueva s/n, 18071 Granada,
Spain}
\address{Departamento de F\'isica At\'omica, Molecular y Nuclear, 
Universidad de Granada, Fuentenueva s/n, 18071 Granada, Spain
}

\author{Francisco de los Santos}
\address{
Instituto Carlos I de F{\'\i}sica Te\'orica y
Computacional, Universidad de Granada, Fuentenueva s/n, 18071 Granada,
Spain}
\address{Departamento de Electromagnetismo y F{\'\i}sica de la
Materia, Universidad de Granada, Fuentenueva s/n, 18071 Granada,
Spain}

\date{\today}

\begin{abstract}

Wave packet revivals and fractional revivals are studied by means of 
a measure of nonclassicality based on the Fisher information. 
In particular, we show that the spreading and the regeneration of initially Gaussian wave packets
in a quantum bouncer and an in the infinite square-well correspond, respectively, to high and low nonclassicality 
values. This result is in accordance with the physical expectations that at a quantum revival 
wave packets almost recover their initial shape and the classical motion revives temporarily afterward.
\end{abstract}
\end{frontmatter}

dlsantos@onsager.ugr.es
+34 958 244 014

\section{Introduction}

Quantum systems may behave classically or quasiclassically under a variety of circumstances and,
in this regard, the transition from quantum to classical mechanics still poses intriguing problems
that attract considerable attention. Of particular interest are systems that display both classical 
and quantum periodic motions, with generally different, incommensurable periods, for in this case the 
interesting question arises as to how the classical periodicity emerges from the quantum one in the 
appropriate limit. For instance, a particle of mass $m$ and energy $E$ in an infinite square-well potential of 
width $L$ initially oscillate with a classical period $T_{\rm cl}=L\sqrt(2m/E)$. The classical oscillations
gradually damp out as the wave packet representing the particle spreads more or less uniformly across the well. 
Quantum mechanically, the  wave function regains exactly its initial form with a {\em revival} period $T_{\rm rev}=4mL^2/\pi \hbar$, 
after which the classical oscillations resume with period $T_{\rm cl}$ again. 
At times that are rational fractions of $T_{\rm rev}$, the wave packet temporarily splits into a number of scaled copies called 
fractional revivals \cite{3,4,rob}.
In the same vein, an object of mass $m$ released from a height $z_0$ and subjected only to gravity, bounces up and down 
against an impenetrable flat surface with a classical period (in suitable units) $T_{\rm cl}= 2 \sqrt{z_0}$,
while the wave function of the corresponding quantum bouncer {\em almost} returns to its initial shape 
after a revival time $T_{\rm rev}= 4z_0^2/\pi$. After a revival has taken place, a new cycle of
quasiclassical behavior and revivals commences again. The fact that in this case the revivals are only approximate
does not make a difference. Revivals and fractional revivals received a great deal of attention over the last
decades. Theoretical progress and experimental observations were  made in atoms and molecules,
and Bose-Einstein condensates \cite{1,2,isotope,primenumbers}. Recently, revivals have been
theoretically investigated in low dimensional systems \cite{17l,18l,19l,25l,26l,nues} and 
have been related to quantum phase transitions \cite{qptrev}.

Identifying the occurrence of wave packet revivals usually 
makes use of the autocorrelation function $A(t)=\langle \Psi(0) | \Psi(t)\rangle$, 
which is the overlap between the initial and the time-evolving wave packet.
Within this approach, the occurrence of revivals and fractional revivals corresponds to, respectively,
the return of $A(t)$ to its initial value of unity and the appearance of relative maxima in $A(t)$. 
Another method to study revival phenomena consists in monitoring the time evolution of the expectation values of some quantities
\cite{rob,sun,don}, and an approach based on a finite difference eigenvalue method has been put forward that allows to predict the
revival times directly  \cite{laserphysics}. 
Recently, information entropy approaches were proposed \cite{rom1} based on the Shannon and R\'enyi entropies, complementary
to the conventional ones. This technique was shown to be superior to analyses based on both the standard variance uncertainty
product \cite{rom3} and the autocorrelation function, inasmuch as it overcomes the difficulty that wave packets reform themselves
at locations that do not coincide with their original ones. A complementary
informational measure is the Fisher information \cite{fisher} which has  attracted
 substantial interest in physics , in particular   in atomic
and molecular physics (see for example 
\cite{dft0,steric,dehesa, dft1,dft2,frieden,frieden1,regi,roman,rsd,ns,rd,romera,shubin,fisah}). 
In this paper we show that the analysis of the wave packet dynamics can be  
carried out using a new tool, 
namely, the {\em nonclassicality} 
$J_{\rm nc}$, defined in terms of the Fisher information 
as we now discuss.

Hall \cite{hall} has recently introduced a measure of nonclassicality,
$J_{\rm nc}$, in terms of the probability 
densities in position and momentum spaces,
$\rho(x)=|\psi(x)|^2$ and $\gamma(p)=|\phi(p)|^2$, respectively.
To be concrete, $J_{\rm nc}\equiv(\hbar/2)\sqrt{I_{\rho} I_{\gamma}}$, where
\begin{equation}
I_{\rho}=4\int \left|\frac{d}{dx}\rho^{1/2}(x)\right|^2 dx, \hspace{0.2cm}
I_{\gamma}=4\int\left|\frac{d}{dp}\gamma^{1/2}(p)\right|^2 dp.
\label{def}
\end{equation} 
Note that $I_\rho$ and $I_\gamma$ are the classical Fisher informations associated
with the probability densities $\rho(x)$ and $\gamma(p)$ \cite{fisher}.
In the next section we shall show that the time evolution of a wave packet exhibiting 
revivals and fractional revivals is initially characterized by a classical
behavior with a low $J_{\rm nc}$ value, followed by a wave packet spreading
with a higher value of $J_{\rm nc}$. In the long time evolution, for times
near $T_{\rm rev}$, the wave packet (approximately) restores its initial form, exhibiting classical periodicity
again accompanied by a decrease in $J_{\rm nc}$. 
In the case of fractional revivals at $t=p T_{\rm rev}/q$, several minipackets emerge whereupon 
a decrease in $J_{\rm nc}$ is expected.

This paper is organized as follows. In Section II we shall consider the Fisher information 
as it applies to revival phenomena. In particular, we show the role of the Fisher information
as a measure of nonclassicality in the dynamics of two model systems that exhibit
fractional revivals:  the so-called quantum
`bouncer', that is a quantum particle bouncing against a hard surface under the
influence of gravity and the infinite square-well. 
Finally, some concluding remarks will be given in the last section.

\section{Fisher information and nonclassicality}

The Fisher information of single particle systems is defined as a
functional of the density function in conjugate spaces by Eq. (\ref{def})
and it has been shown to be a measure of nonclassicality \cite{hall}. Following Hall \cite{hall}
and Mosna et al. \cite{mosna}, the Fisher information in position space can be expressed as
$I_{\rho} = (4/\hbar)^2 (\langle P^2\rangle_{\psi}-\langle P^2_{\rm cl}\rangle_{\psi})$  
where $P$ denotes the momentum operator and $P_{\rm cl}$ is a (state-dependent) classical momentum
operator defined by
\begin{equation}
P_{\rm cl} \Psi(x)=\frac{\hbar}{2i}\left(
\frac{\psi^{\prime}(x)}{\psi(x)}-\frac{\psi^{*\prime}(x)}{\psi^{*}(x)} \right)
=\hbar\Big(\arg \psi(x)\Big)^{\prime}.
\end{equation}
Hence, it is natural to separate the momentum operator in a classical ($P_{\rm cl}$) and
nonclassical ($P_{\rm nc}$) contribution with $P_{\rm nc}\equiv P-P_{\rm cl}$. 
The definition of the classical momentum observable is supported by the facts that
$\rho$ satisfies the classical continuity equation and that the expectation values of $P$ and $P_{\rm cl}$ are equal for
all wave functions $\langle P\rangle_{\psi}=\langle P_{\rm cl}\rangle_{\psi}$ \cite{hall}.
The conjugate equality that relates the momentum Fisher information and the nonclassicality of the position operator
can be obtained analogously,
$
I_{\gamma}=(4/\hbar)^2(\langle X^2\rangle_{\phi}-\langle X^{2}_{\rm cl}\rangle_{\phi}) 
$ 
Finally, Hall introduced a measure of joint nonclassicality $J_{\rm nc}$ for a quantum state
as 
\begin{equation}
J_{\rm nc}\equiv \frac{\hbar}{2}I^{1/2}_{\rho}I^{1/2}_{\gamma}.
\end{equation}
It follows that $J_{\rm nc}=1$ for Gaussian distributions. For instance, the evolution of Gaussian wave
packets in a harmonic oscillator follows a periodic motion in
accordance with classical expectations \cite{rob}, and $J_{\rm nc}=1$ for all times. 
For mixed states, $J_{\rm nc}$ can be arbitrarily small while for pure
states Hall found  $J_{\rm nc}\geq |1+(i/\hbar)\langle[P_{\rm cl},X_{\rm cl}]\rangle_{\psi}|$ \cite{hall}.

\subsection{Quantum bouncer}

Consider an object of mass $m$  bouncing against a hard surface subjected only
to the influence of the gravitational force directed downward along the $z$ axis, that is, 
a particle in a potential $V(z)=mgz$, if  $z>0$ and  $V(z)=+\infty$ otherwise.
Gravitational quantum bouncers have been recently realized using neutrons \cite{qb1} and atomic clouds \cite{qb2}. 
Their revival behavior has been discussed in \cite{don,gea} and  an entropy-based approach was presented in
\cite{rom1,rom3}.

The time-dependent wave function for a localized quantum wave packet is expanded as
a one-dimensional superposition of energy eigenstates as 
\begin{equation}
\psi(x,t)=\sum_n a_n u_n(x) e^{-i E_n t/\hbar}.
\label{evolutionx}
\end{equation}
 The eigenfunctions and eigenvalues are given by \cite{gea}
\begin{equation}
E^{\prime}_n=z_n; \quad u_n(z^{\prime})= {\cal N}_n
{\rm Ai} (z^{\prime}-z_n); \quad n=1,2,3,\ldots
\end{equation} 
where 
$l_g=\left(\hbar/2gm^2\right)^{1/3}$ is a characteristic gravitational length,
$z^{\prime}=z/l_g$, $E^{\prime}=E/mgl_g$,  Ai$(z)$ is the Airy function,
$-z_n$ denotes its zeros, and ${\cal N}_n=|{\rm Ai}'(-z_n)|$ is the $u_n(z^\prime)$ normalization factor. 
Accurate analytic approximations for $z_n$ exist \cite{gea},
\begin{equation}
z_n \simeq \frac{3\pi}{2}\left[ n -\frac{1}{4}\right]^{2/3}.
\end{equation}
Consider now an initial Gaussian wave packet 
localized at a height $z_0$ above the surface, with a width $\sigma$ and a
momentum $p_0$ (in the remainder of this paper the primes on the energy and position
symbols will be dropped)
\begin{equation}
\psi(z,0)=\frac{1}{\sqrt{\sigma\hbar\sqrt{\pi}}}e^{-(z-z_0)^2/2\sigma^2\hbar^2}
e^{ip_0(z-z_0)/\hbar}.
\label{paqueteinicialqb}
\end{equation}
If the lower bound of the integral is extended to $-\infty$, 
the associated coefficients of the time-dependent wave function for $p_0=0$ 
can be obtained analytically as \cite{gea}
\begin{eqnarray}
C_n &\simeq& {\cal N}_n \left( \frac{2}{\pi \sigma^2}\right)^{1/4} \int_{-\infty}^{\infty}
{\rm Ai}(z-z_n) e^{-(z-z_0)^2/\sigma^2} dz  \nonumber \\
 &=& {\cal N}_n\left( \frac{2}{\pi \sigma^2}\right)^{1/4} \sqrt{\pi} \sigma 
\exp\left[\frac{\sigma^2}{4}\left(z_0-z_n+\frac{\sigma^4}{24}\right) \right] \nonumber \\
&&\times {\rm Ai}\left(z_0-z_n+\frac{\sigma^4}{16}\right).
\end{eqnarray}

The important time scales of a wave packet's time evolution are in the
coefficients of the Taylor series (see, for instance \cite{3,4,rob}) of the
energy spectrum $E_n$ around the level the wave packet is centered around, let us say  $\bar{n}$:
\begin{equation}
E_{\bar{n}}= E_n+ 2\pi\hbar \left(\frac{(n-\bar{n})}{T_{\rm cl}}+ \frac{(n-\bar{n})^2}{T_{\rm rev}}+ \cdots\right).
\end{equation}
The classical period and the revival time 
can be calculated to obtain $T_{\rm cl}=2\sqrt{z_0}$ and $T_{\rm rev}=4 z_0^2/\pi$, respectively \cite{gea}. 
The temporal  evolution of the wave packet in momentum-space was obtained numerically by the fast Fourier 
transform method.

\begin{figure}
\psfig{width=7cm, angle=270, file=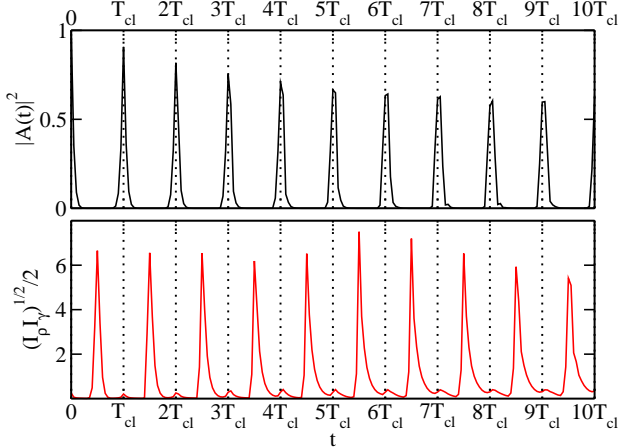}
\caption{Autocorrelation function (top panel) and nonclassicality (bottom panel)
for the early time evolution of a quantum bouncer with $z_0=100$, $\sigma=1$, and $p_0=0$. 
The classical time is $T_{\rm cl} =20$ and the classical periods are denoted by vertical 
dotted lines.}
\label{f4}
\end{figure}

\begin{figure}
\psfig{width=7cm, angle=270, file=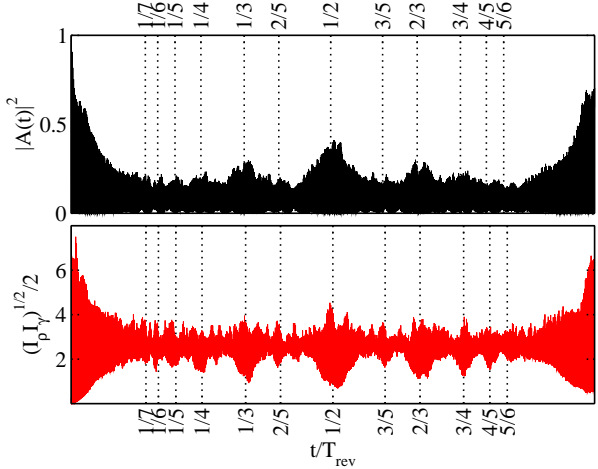}
\caption{Autocorrelation function (top panel) and nonclassicality (bottom panel)
for a quantum bouncer with $z_0=100$, $\sigma=1$, and $P_0=0$. The revival time 
is $T_R \approx 12~732.4$ and the main fractional revivals are denoted by vertical 
dotted lines.}
\label{f5}
\end{figure}

We have computed the temporal evolution of the autocorrelation function and 
the nonclassicality $J_{\rm nc}$ for the initial conditions $z_0=100$,
$\sigma=1$ and $p_0=0$. Figure \ref{f4} displays the early time evolution of both 
quantities and the location of the main fractional revivals. The top panel shows 
how the autocorrelation function initially follows the first classical periods of motion.
The nonclassicality, $J_{\rm nc}$, describes precisely this same behavior, with peaks at the 
wave packet's collapses and minima at the multiples of the classical period.  
In the long time limit, the wave packet eventually spreads out and collapses, only 
to reform at multiples of the revival time. In between, fractional revivals take place.
All this is reflected in the maxima and relative maxima of $|A(t)|^2$ as shown in the top panel
of Fig. \ref{f5}. The alternative description in terms of $J_{\rm nc}$ is shown in the 
bottom panel of Fig. \ref{f5}. The slightly nonclassical behavior or, equivalently, the quasiclassical 
behavior that takes place at full and fractional revivals is described by, respectively, the minima and 
the relative minima of $J_{\rm nc}$. Finally, notice that the occurrence of fractional revivals at, for 
instance,  $t=4/5 T_{\rm rev}$ or $t=5/6 T_{\rm rev}$  is clearly better indicated by $J_{\rm nc}$ than by $|A(t)|^2$.

\subsection{Infinite square-well system}

The one-dimensional infinite square-well potential confines a particle of mass
$m$ to a box of width $L$ and is described by
$V(x)=0$ for $0<x<L$ and $V(x)=+\infty$ otherwise.

The normalized eigenstates and  the corresponding eigenvalues are given by,
\begin{equation}
u_n(x)=\sqrt{\frac{2}{L}} \sin{\left(\frac{n\pi
x}{L}\right)},\quad\quad E_n=n^2\frac{\hbar^2\pi^2}{2 m L^2}.
\end{equation}
The classical period and the revival time can be computed as
$T_{\rm cl}=2 m L^2/\hbar \pi n$ and $T_{\rm rev}=4mL^2/\hbar \pi$, respectively.
Note that due to the fact that in this particular case the quantized energy levels are exactly
quadratic in $n$, there is no super-revival (or higher-order effects in the
evolution), nor does the revival period depend on the mean energy level $\bar{n}$.
The occurrence of revivals and fractional revivals can now be illustrated
by simply taking $t=kT_{\rm rev}$ in the expansion (\ref{evolutionx}),
which immediately entails $\psi(x,t=kT_{\rm rev})=\psi(x,t=0)$,
with $k$ an integer \cite{4}. Therefore, the time evolution of the 
infinite square-well is periodic with period $T_{\rm rev}$, and this period is an exact 
revival time too. It is also easy to see by direct substitution in (\ref{evolutionx})
that $\psi(L-x,T_{\rm rev}/2)=-\psi(-x,0)$, so at time $t=T_{\rm rev}/2$ a copy of the initial 
state reforms itself, reflected around the center of the well \cite{4}.

We shall consider initial Gaussian wave packets centered at a position $x_0$, with
a momentum $p_0$ and a variance $\sigma$,
\begin{equation}
\psi(x,0)=\frac{1}{\sqrt{\sigma\hbar\sqrt{\pi}}}e^{-(x-x_0)^2/2\sigma^2\hbar^2}
e^{ip_0(x-x_0)/\hbar}.
\label{paqueteinicial}
\end{equation}
Assuming that the integration region can be extended to the whole real axis 
due to the exponential functional form of (\ref{paqueteinicial}) \cite{rob},
the expansion coefficients can be approximated with high accuracy by the analytic 
expression
\begin{eqnarray}
a_n\approx
2\sqrt{\frac{b\sqrt{\pi}}{L}}
e^{-b^2[p_0^2+(\frac{n\pi\hbar}{L})^2]} 
\sin\left[ \frac{n\pi}{L}
\left(x_0+\frac{ib^2p_0}{\hbar} \right)\right],
\end{eqnarray}
where $b=\sigma \hbar$, $p_n=n\pi\hbar/L$ and  $p_0=n_0\pi\hbar/L$.
To calculate the corresponding time dependent, momentum wave function
we use the Fourier transform of the equation (\ref{paqueteinicial}), 
and the momentum-space normalized eigenstates 
\begin{equation}
\phi_n(p)=\sqrt{\frac{\hbar}{\pi L}}
\frac{p_n}{p^2-p^2_n}\bigg[(-1)^n e^{ipL/\hbar}-1\bigg].
\end{equation}
The initial wave packet in momentum space is then given 
by the Fourier transform of the equation (\ref{paqueteinicial}), 
which leads again to a Gaussian expression,
\begin{equation}
 \Phi(p,0)=\sqrt{\frac{\sigma}{\sqrt{\pi}}}
e^{-\sigma^2(p-p_0)^2/2}e^{-ip x_0/\hbar}.
\end{equation}
Without loss of generality, we shall henceforth take $2m=\hbar=L=1$, 
and $\sigma=1/\sqrt{200}$ for the initial wave packet.

\begin{figure}
\psfig{width=6cm, angle=270,file=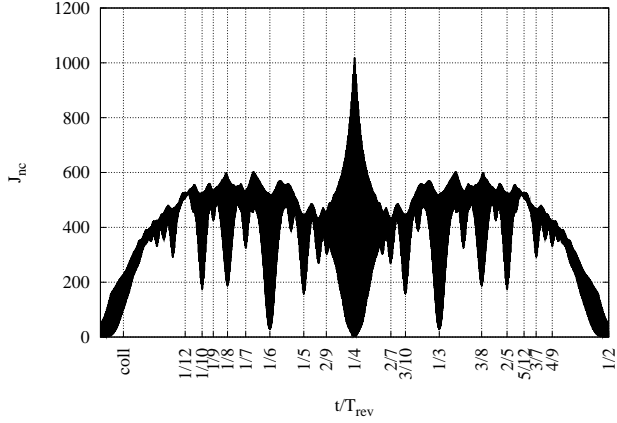}
\caption{Nonclassicality for the time evolution of wave packet in 
an infinite square-well with parameters $x_0=0.5$ and
$p_0=400\pi$.  The main fractional revivals are indicated by 
vertical dotted-lines.}
\label{suma400}
\end{figure}

We have computed the measure of nonclassicality  $J_{\rm nc}$ and the
autocorrelation function for an initial wave packet with $x_0=0.5$ and
$p_0=400\pi$. The results are shown in Fig. \ref{suma400}. 
At early times, 
the Gaussian wave packet evolves quasiclassically but in a few periods
the quantum and classical wave packet trajectories start moving apart, 
the classical component of the wave function being defined as 
$\psi_{\rm cl}(x,t)=\sum_n a_n u_n(x) e^{-i 2\pi n t/T_{\rm cl}}$ \cite{rob}. 
For long time scales, a large amplitude modulation with a set of  
relative minima is superimposed on the quasiperiodic oscillations (Fig. \ref{suma400}). 
In this long-time regime, the wave packet initially spreads and delocalizes while undergoing a sequence 
of fractional revivals with the creation of correlated sets of sub-packets 
located along the classical orbit, each of them similar to the initial one
so that the nonclassicality  reaches a relative minimum, 
the most important fractional revivals are denoted by vertical dashed-lines. 
Notice that at one half of the revival period there is an absolute 
minimum which corresponds to a single rejected copy of the initial wave function.

\section{Summary}

To summarize, we have shown how the measure of nonclassicality $J_{\rm nc}$ accounts for  
the regeneration of initially well localized wave packets during their time evolution. 
In particular, $J_{\rm nc}$ shows the spreading (high nonclassicality values) and the resuming
of the classical periodic motion (low nonclassicality values) of wave packets in two example 
cases, namely, the quantum bouncer and the infinite square-well. 
This approach overcome the generic difficulty that the autocorrelation function misses to detect
some fractional revivals because the mini-packets emerge at arbitrary positions that do not coincide with
the original one. This appears to be a common advantage of all the entropy-based approaches. 

This work was supported by the Spanish Projects No.
MICINN FIS2009-08451, No. FQM-02725 (Junta de Andaluc\'ia), 
 and No. MICINN FIS2011-24149.

\end{document}